\documentclass[prb,showpacs,floatfix,twocolumn]{revtex4}
\usepackage{epsfig}
\begin{document} 
\title{The optical conductivity of half-filled Hubbard ladders}
\author{J. Hopkinson and K. Le Hur}
\affiliation{D\'epartement de Physique and CERPEMA,
 Universit\'e de Sherbrooke, Sherbrooke, Qu\'ebec, Canada, J1K 2R1}

\newcommand{\br}{{\bf r}}
\newcommand{\ovl}{\overline}
\newcommand{\hw}{\hbar\omega}
\newcommand{\mybeginwide}{
    \end{multicols}\widetext
    \vspace*{-0.2truein}\noindent
    \hrulefill\hspace*{3.6truein}
}
\newcommand{\myendwide}{
    \hspace*{3.6truein}\noindent\hrulefill
    \begin{multicols}{2}\narrowtext\noindent
}
 
\date{\today} 
\begin{abstract}
We investigate the optical conductivity of half-filled N-leg Hubbard ladders
far into the ``deconfinement'' limit (i.e., weak Hubbard interaction and 
relatively strong interchain hopping). The N-leg Hubbard ladder
is equivalent to an N-band model with velocities obeying 
$v_1=v_N<v_2=v_{N-1}<...$. When N is not too large $(N=3,4...)$, the band 
pairs $(i,N+1-i)$ successively flow to the D-Mott state 
leading to a cascade of charge and spin 
gaps [U. Ledermann, K. Le Hur, and T. M. Rice, Phys. Rev. B {\bf{62}}, 16383 
(2000)], and to the 
progressive closing of the two-dimensional (2D) Fermi surface (FS). 
The optical conductivity at finite temperatures can then exhibit
coexistence between a prominent Drude peak and a 
high-frequency preformed pair continuum, split by sharp excitonic peaks 
arising due to an approximate SO(8) symmetry. 
For very large (but finite) N, all neighboring bands interact on the 
2D FS, leading to a low-temperature 2D Mott crossover 
accompanied by
a Spin Density Wave (SDW) instability (similar to the 2D case).  In this limit 
the optical conductivity exhibits a unique charge gap -excitations above 
the gap being bound hole-pairs- and that the exciton
features vanish.  These results could help to explain
the optical conductivity of 2D systems at and 
close to half-filling, an example of which is the pseudogap phase
of high-$T_c$ cuprates.

\end{abstract}

\pacs{71.10.Pm;71.30.+h;72.10.-d}
\maketitle

{\centering{
{I. INTRODUCTION\\}}}

\vskip1pc

Understanding the nature of the interplay between {\it strong} interactions 
(generally emerging at the
stoichiometric electron density due to the possibility of umklapp scattering)
and {\it dimensionality} is an important
open matter relevant to a large class of materials including the cuprate
superconductors and quasi one-dimensional (1D) organic conductors.
The latter can be viewed as possible realizations of 
coupled one-dimensional Hubbard chains at half-filling 
where electrons can hop from chain 
to chain (although the quarter-filled chains are weakly dimerized allowing half-filled umklapps, there is increasing evidence that quarter-filled umklapp scattering dominates the physics)\cite{Claude1}. In particular, the Fabre salt
family (TMTTF)$_2$X (X = ClO$_4$, Br and PF$_6$) displays 
insulating behavior
at ambient pressure up to quite high temperatures. This reflects the 
presence of strong umklapp scattering (interaction, $U$) along the 
chains due to the commensurate
filling resulting in a 1D Mott transition; essentially one has virtually 
uncoupled insulating chains 
({\bf confinement})\cite{Thierry,Karyn1,Tsvelik}. 

By contrast, substitution of Se for S to create the
Bechgaard salt (TMTSF)$_2$PF$_6$, increases 
the hopping between chains (t$_\perp$) or alternately decreases the dimerization sufficiently to 
delocalize particles in the transverse direction 
(inducing {\bf deconfinement}). The
system thus exhibits a crossover to a regime of metallic planes (as t$_\perp$ 
scales to strong coupling faster than $U$).
Experimentally
this dimensional crossover takes place around $100K$ in (TMTSF)$_2$PF$_6$; This
is manifested by a change from $T$ 
(single-chain Luttinger liquid) to $T^2$ (Fermi liquid) behavior
of the dc transport along the chain axis{\cite{jerome}}. However, such compounds 
are 
located quite close to the confinement transition such that the optical 
conductivity of (TMTSF)$_2$PF$_6$ 
exhibits many features in common with the single-chain Mott insulator\cite{Vescoli} 
(TMTTF)$_2$PF$_6$. Indeed, only $1\%$ of the spectral weight\cite{Vescoli} 
contributes to the dc (Fermi liquid) 
transport. Some theoretical attempts to reproduce
this result (which bears a superficial resemblance to our 
half-filled chains for small N (see Fig. 1)),  have been performed recently
using Dynamical Mean Field Theory (DMFT)\cite{Thierry2} and RPA\cite{Tsvelik2}.

Here, we are additionally motivated to tackle the physics of 2D Cu-O planes
of cuprate materials at and close to half-filling where Mott physics
(``Mottness''\cite{phillips}) is also ubiquitous\cite{Maurice};
We systematically investigate the behavior of
the optical conductivity for half-filled N-leg 
Hubbard ladders (=coupled chains) {\it far} into the ``deconfinement'' regime
(the bare transverse hopping amplitude is almost equal to the longitudinal
hopping amplitude). For weak on-site repulsion, we can make use of our 
previous band approach 
for half-filled N-leg Hubbard ladders\cite{Karyn2}. This allows 
us to properly take into account the renormalization of the 
different coupling channels (umklapp, Cooper,...) 
and also to provide a rigorous study of 
the ground state. The N-leg Hubbard ladder
is equivalent to an N-band model where at half-filling the Fermi velocities
obey $v_1=v_N<v_2=v_{N-1}<...$. 

For small N $(N=3,4...)$, this model results in a cascade of 
energy scales\cite{Karyn2}, where band pairs $(i,N+1-i)$ successively flow to 
the D-Mott state which possesses an enlarged
SO(8) symmetry; This is reminiscent of the two-leg ladder behavior\cite{Lin}. 
This leads to a complex behavior of the optical
conductivity: a Drude peak and a high-frequency
particle-hole continuum coexist, with sharp exciton peaks appearing
below the continuum reflecting the underlying  SO(8) symmetry.  

By increasing 
considerably the number of chains, ``four-band'' couplings (Fig.~2) 
become relevant 
on the 2D Fermi surface.  At sufficiently 
low temperatures, a 2D insulating transition with strongly enhanced SDW 
correlations arises\cite{Urs}. {\it The 
low-energy physics already converges to that
of the purely 2D Hubbard model}\cite{Zanchi,Maurice}. We rigorously 
establish that charge excitations above the Mott gap
consist of bound hole-pairs (preformed pairs) 
and that the excitonic peaks cannot
survive in this limit. This could be relevant to an understanding of 
optical conductivity measurements on the high-$T_c$ cuprates at and close to 
half-filling\cite{optical}; Indeed, umklapp scattering should not be
ignored in the underdoped regime of the cuprates--features seen above
$\sim 2eV$ and the transfer of spectral weight as a function of doping or 
temperature (T) reflect this importance.

\vskip1pc

{\centering
{II. THE MODEL }\\}

\vskip1pc

Our starting point is the N-leg Hubbard ladder model,
\begin{eqnarray}
  H_{Kin}&=&-t\sum_{x,i,s} d_{is}^{\dagger}(x+a)d_{is}(x)
    +{\mathrm h.c.}
  \nonumber\\
  &&-t_{\perp}\sum_{x,i,s}d_{i+1s}^{\dagger}(x)d_{is}(x)
    +{\mathrm h.c.};
\end{eqnarray}
$t$ and $t_{\perp}$ denote the hopping matrix elements along and 
perpendicular to the chains, $a$ is a lattice step, and $d_{is}(x)$ annihilates an electron 
with spin $s$ on chain $i$ at the rung $x$.
The interaction term reads
\begin{equation}
  H_{\rm Int}=U\sum_{i,x}d_{i\uparrow}^{\dagger}(x)d_{i\uparrow}(x)
    d_{i\downarrow}^{\dagger}(x)d_{i\downarrow}(x),
\end{equation}
where $U$ is the on-site Hubbard repulsion.
Again, here we are 
investigating the perfect ``deconfinement'' regime which implies that
the perpendicular hopping $t_{\perp}$ is sufficiently large to 
deconfine {\it all} the electrons in the transverse direction. More precisely, 
below, we consider the weak-interaction limit $0<U\ll (t,t_{\perp})$ where
tractable (and rigorous) calculations are indeed possible\cite{Karyn2}.  
It is convenient
to use the {\it band picture} where the kinetic part simply takes a diagonal
form:
\begin{equation}
  H_{Kin}=\sum_{i=1...N,s}\int dk\ \epsilon_{i}(k)\Psi_{is}^{\dagger}(k)
\Psi_{is}(k);
\end{equation}
$\Psi_{is}^{\dagger}$ and $\Psi_{is}$ are the creation
and annihilation operators for the band~$i$ and
\begin{equation}
  \epsilon_{i}(k)=-2t\cos(ka)-2t_{\perp}\cos(k_{\perp i}a).
\end{equation}
The transverse (Fermi) momenta here simply obey $k_{\perp i}=\pm \pi i/(a(N+1))$
and at half-filling the longitudinal momenta are exactly determined by 
$\epsilon_{i}(k_{Fi})=0$. The resulting Fermi velocities 
$v_{i}=\frac{2ta}{\hbar}\sin(k_{Fi}a)$ then take the form
\begin{equation}
  v_{i}=v_{\bar{\imath}}=\frac{2a}{\hbar}\sqrt{t^{2}-
\left\{t_{\perp}\cos[\pi i/(N+1)]\right\}^{2}},
\end{equation}
where $\bar{\imath}=N+1-i$. The transformation from chain to band 
(with open boundary conditions) reads\cite{Lin2}:
\begin{equation}\label{fermion}
d_{is}=\sum_m \sqrt{\frac{2}{N+1}}\sin \hbox{\huge{(}} \frac{\pi m i}{N+1}
\hbox{\huge{)}}\Psi_{ms}.
\end{equation}
 For more details on the
method at half-filling, see Ref.~\onlinecite{Karyn2}. 

Two different limits arise.  When $N$ is {\it not too
large}, careful investigation of the coupled Renormalization Group 
equations (RGEs) of the  
interacting N-band
problem yields a hierarchy of energy scales,  
$T_i\sim t e^{-\alpha v_i\hbar/(aU)}$ ($\alpha$ is a constant
parameter of the order of 1), $T_1>T_2>...>
T_r$ --for N even $r=N/2$ and for N odd $r=(N+1)/2$, where band pairs
$(i,\bar{\imath})$ decouple from all other bands and freeze out.  This 
decoupling, due to band pair umklapp terms satisfying 
$(k_{Fi}+k_{F\bar{\imath}})=\pi/a$, opens both a charge and spin 
gap. The effective {\it single-particle} gaps are approximately 
given by $\Delta_j=T_j$. More precisely, the couplings of the band pairs
$(i,\bar{\imath})$ scale towards the two-leg ladder fixed point with 
SO(8) symmetry\cite{Lin}. To compute the optical conductivity in this 
limit, it will
be crucial to exploit this enlarged symmetry. 
When $N$ becomes {\it large} (but finite), our method
is still rigorous provided the energy difference between two neighboring bands
is much larger than the largest relevant 
energy gap in the problem, i.e., $t e^{-t/U}$.
This imposes the condition\cite{Lin2} $U<t/\ln N,t_{\perp}/\ln N$. Here, 
coupling
between \underline{neighboring} bands 
leads to a 2D-like SDW instability (crossover) --opening a Mott gap-- 
at the temperature\cite{Urs}:
\begin{equation}\label{sdw}
T_{sdw}\sim t e^{-\frac{N}{U\sum_{i=1}^{N/2} \frac{a\hbar}{v_i}}}.
\end{equation}
Since we are considering relatively large bare values of $t_{\perp}$ (i.e., 
perfect nesting on the 2D FS), the
resulting charge gap $\Delta\sim T_{sdw}$ 
is homogeneous on the 2D Fermi surface. When substantially 
decreasing $t_{\perp}$
(approaching the confinement transition), we expect that 
the physics would have 
a larger
dependence on the transverse momenta\cite{Claude2,Tsvelik2} $k_{\perp i}$. 
Finally, ground-state charge properties still resemble those of the small-N 
(two-band) limit.

\vskip1pc
{\centering
{III. SMALL N: MAPPING TO SO(8) }\\}

\vskip1pc

For small N, the low-energy Hamiltonian is then the 
sum of $N/2$ [(N-1)/2 for N odd] two-leg ladder Hamiltonians 
$H_{i,\bar{\imath}}$
corresponding
to the band pairs $(i,\bar{\imath})$ plus the Hamiltonian of a single chain 
for N odd $H_s$\cite{Karyn2}.

 Let us consider the case $N=3$ for simplicity. Here,
the relevant 
couplings of bands 1 and 3 flow towards the universal ratios of the 
two-leg ladder,
 and therefore the low-energy Hamiltonian $H_{1,3}$ can be rewritten in
terms of an SO(8) Gross-Neveu model (as shown previously by 
Lin, Balents, and Fisher for the two-band model\cite{Lin})
\begin{equation}
H_{1,3}= \int dx\ \biggl\{-i v_1\sum_{a=1}^4
{\Psi}^{\dagger}_a\tau^z\partial_x\Psi_a-
g\hbox{\large{(}}\sum_{a=1}^4\Psi_a^{\dagger}\tau^y\Psi_a\hbox{\large{)}}^2\biggr\}.
\end{equation}
We have implicitly introduced operators $\Psi_{R/L a}$ for right and left 
moving fermions and the Pauli matrices $\tau^i$ act on right and 
left sectors.
Even though the four Dirac fermions above are related to 
the {\it charge} and {\it spin} degrees of the two bands, they cannot be simply
identifed as the original (real) fermions on each band; for a complete map, see
Ref.~\onlinecite{Lin}. The low-energy physics
emerging from bands 1 and 3 depends on a single coupling
$g$ (whose bare value is 
of the order of $U$). It is straightforward to show that the interacting
part of $H_{1,3}$ both opens a charge and spin gap equal to
$2\Delta_1=4g<{\Psi}^{\dagger}_a\tau^y\Psi_a>\ \sim 2T_1$. The 
ground state corresponds to the ``D-Mott state'': a Mott insulator having
short-range pairing correlations with approxi- 

\begin{figure}[ht]
\centerline{\epsfig{file=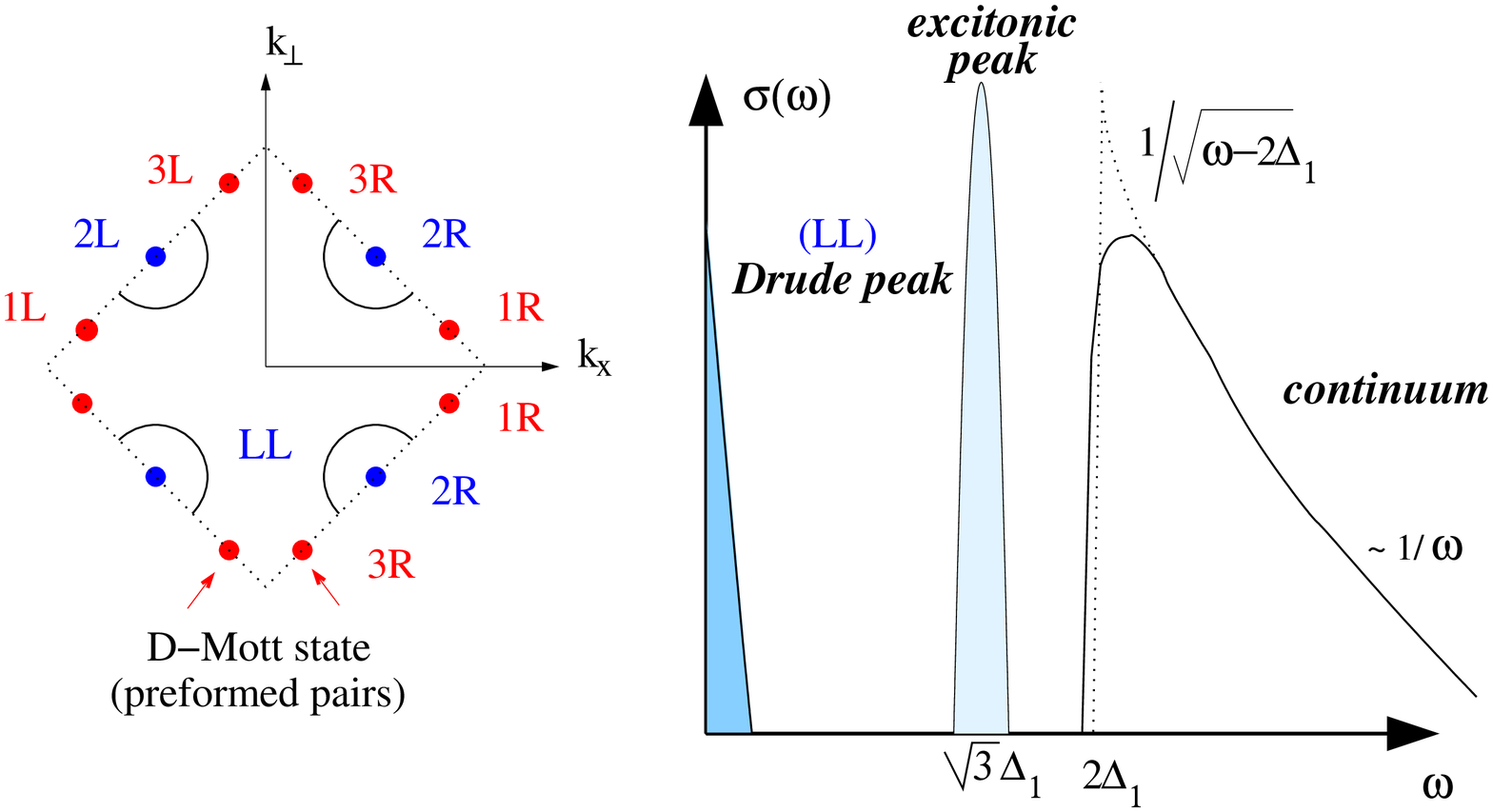,angle=0.0,height=4.7cm,width=
8.95cm}}
\caption{Optical conductivity of the three-band model for
$T_2\ll T\ll T_1$. The Fermi surface (FS) (drawn for $t_{\perp}=t$)
exhibits a blatant truncation, 
meaning that bands 1 and 3 form the D-Mott state opening
a gap whereas the band 2 is still metallic (and described by a 
Luttinger theory). The optical conductivity then exhibits a Drude peak
(due to band 2), a high-energy continuum (treated here at the mean field level) above the charge gap $2\Delta_1$
and a sharp exciton peak due
to the SO(8) symmetry of the D-Mott state.}
\end{figure}

\hskip -0.3cm -mate d-wave symmetry\cite{Lin}. 
Additionally, 
fluctuations of the gap around its vacuum value can generate
{\it attractive} interactions between the fermions $\Psi_a$ 
leading to the formation of {\it excitons}, whose mass satisfies
$M_1=\sqrt{3}\Delta_1$ (i.e. is smaller than the charge gap of the (localized)
preformed-pair continuum\cite{Lin}). Unlike 
the
single-chain\cite{Venky}, here excitons already appear for purely 
on-site repulsion.

Band 2 is described by a single chain at half-filling $(\mu =0)$, which 
is embodied by gapless spinons (spin-1/2 excitations) and the following 
charge Hamiltonian:
\begin{eqnarray}
H_s^{\rho}=\int dx\biggl\{\frac{u_{\rho}}{2}
    \left[\frac{1}{K_{\rho}}(\partial_{x}\Phi_{\rho})^{2}+K_{\rho}
(\partial_{x}\Theta_{\rho})^{2}\right]\biggr.
  \nonumber\\
    \biggl.-\frac{g_{\rho}}{(\pi a)^2}\cos(\sqrt{8\pi}\Phi_{\rho})-\mu\partial_{x}\Phi_{\rho}\biggr\};
\end{eqnarray}
$\partial_{x}\Phi_{\rho}$ describes charge density fluctuations and
$\partial_{x}\Theta_{\rho}$ current excitations. $g_{\rho}$ is of the order of $U$ and
$u_{\rho}$ is almost equal to $v_2$. It is immediate to check that 
$H_2$ induces a charge gap $2\Delta_2\sim 2T_2<2T_1$.
Let us now compute the optical conductivity of the three-leg ladder. 

Using
formula (\ref{fermion}), the 
electrical current can be written in the chain basis as:
\begin{eqnarray}
\hskip -0.2cm 
J(x)&=&\frac{e}{2i\sqrt{\pi}}\sum_{is}\frac{1}{\delta t_i}\hbox{\Large{(}}d_{is}^{\dagger}(x+a)d_{is}(x) -
d_{is}^{\dagger}(x)d_{is}(x+a)\hbox{\Large{)}}\nonumber
\\ \nonumber
&=&\frac{e}{2i\sqrt{\pi}}\hbox{\Large{(}} \frac{v_{D1}}{a}(\Psi_{2s}^{\dagger}(x+a)\Psi_{2s}(x)-
\Psi_{2s}^{\dagger}(x)\Psi_{2s}(x+a)) \\ \nonumber
&&\hskip-1pc +\frac{v_{D1} + v_{D2}}{2a}(\Psi_{1s}^{\dagger}(x+a)\Psi_{1s}(x)-
\Psi_{1s}^{\dagger}(x)\Psi_{1s}(x+a) \\ \nonumber
&&+ \Psi_{3s}^{\dagger}(x+a)\Psi_{3s}(x)-
\Psi_{3s}^{\dagger}(x)\Psi_{3s}(x+a) ) \\ \nonumber
&& \hskip-1pc +\frac{v_{D1}-v_{D2}}{2a} (\Psi_{1s}^{\dagger}(x+a)\Psi_{3s}(x)+ \Psi_{3s}^{\dagger}(x+a)\Psi_{1s}(x) \\ 
&&-
\Psi_{1s}^{\dagger}(x)\Psi_{3s}(x+a) -
\Psi_{3s}^{\dagger}(x)\Psi_{1s}(x+a)) \hbox{\Large{)}},
\end{eqnarray}
where we have summed up over the contribution coming from each spin $s$ and
each leg $i =1,2,3$ of the ladder.  Symmetry of the ladder dictates that the drift velocity ($v_{D1}$) in chains 1 and 3 be equal, although not necessarily the same as in chain 2 ($v_{D2}$), and the time interval for a hop of one lattice spacing, $a$,  has been taken as $\delta t_i$ = $\frac{a}{v_{Di}}$.  Note that were these velocities equal, the coefficients of the first two terms in the band basis would be identical while the third would disappear leaving a straight sum over bands. Indeed, the last term is found to contain an oscillatory contribution in $x$, which, when inserted into Eq. 12 vanishes.  In linear response, the 
optical conductivity (at $k=0$) takes the standard form
\begin{equation}
{\Re e}\ \sigma(\omega)={\Im m}\hbox{\Huge{(}} \frac{\Pi(\omega)}{\omega}
\hbox{\Huge{)}},
\end{equation}
the current-current correlator being
\begin{equation}
\Pi(\omega)=\frac{1}{\hbar}\int dx d\tau e^{i\omega_n\tau}\
\langle T_{\tau} J(x,\tau)J(0,0) \rangle_{\omega_n\rightarrow -i\omega +\delta}.
\end{equation}

In the three-leg ladder, we find the following current 
\begin{eqnarray}
J&=&\frac{\sqrt{2}}{\pi}e \hbox{\large{(}}
v_{D1}\partial_x\Theta_{\rho}
+\frac{v_{D1} + v_{D2}}{2}\sqrt{2} \sin(k_{F1}a)\partial_x \Theta_{\rho+}\hbox{\large{)}} \nonumber \\ &=& J_{1 band} + J_{2 band}, 
\end{eqnarray}
where one can re-express the 2-band contribution in terms of Majorana fields (real chiral fermions $\psi_{R/L a} = \eta_{R/L 2a-1} + i\eta_{R/L 2a}$) as
\begin{eqnarray}
\hskip-2pc J_{2 band} &=& - \frac{(v_{D1} + v_{D2})e}{\sqrt{2\pi}a}\sin (k_{F1}a)\Psi_1^{\dagger}\tau^z\Psi_1 \nonumber \\
&& \hskip-4pc = -\frac{(v_{D1} + v_{D2})ie\sqrt{2}}{\sqrt{\pi}a}\sin(k_{F1}a)(\eta_{R1}\eta_{R2} - \eta_{L1}\eta_{L2}).
\end{eqnarray}
Using Eq. 12, the contribution of these terms to the conductivity is,
\begin{eqnarray}
{\Re e}\ \sigma_{2 band}(\omega) &&\hskip-1pc= \frac{-e^2(v_{D1} + v_{D2})^2\sin^2(k_{F1}a)}{\pi a^2\hbar\omega} {\Im m}\hbox{\Large{\{}}\int{dx}\nonumber \\ &&\hskip-6pc\int{d\tau} e^{i\omega \tau}\sum_{P={R,L}}<T_{\tau}\eta_{P1}\eta_{P2}(x,\tau)\eta_{P1}\eta_{P2}(0,0)>\hbox{\Large{\}}}.
\end{eqnarray}
As pointed out in Ref. 12, since the two-band model is equivalent to the SO(8) Gross-Neveu model whose excitation spectrum is known, below the D-Mott gap mass $M_1$ bound-states of charge $\pm 2e$ fundamental fermions can form.  
The latter factor of Eq. 15 bears resemblance to the ($\vec{k}=0$) Green's function of a localized particle of mass $M_1$, so that
\begin{eqnarray}
&&\hskip-2pc-\int{dx}\int{d\tau}e^{i0x}e^{i\omega_n\tau}<T_{\tau}\eta_1\eta_2(x,\tau)(\eta_1\eta_2(0,0))> \nonumber \\ &\approx& -\int{dx}\int{d\tau}e^{i0x}e^{i\omega_n\tau}A e^{-\frac{|x|}{\xi}}e^{-\frac{M_1 \tau}{\hbar}} \nonumber\\  &\approx& \frac{A(\xi - \xi e^{-\frac{L}{\xi}})}{i\omega_n - \frac{M_1}{\hbar}}, 
\end{eqnarray} 
where we have introduced a normalization constant, A, the length of the system, L, which we can take to be infinite relative to the correlation length, $\xi$ as we have power law correlations in this system.  
This means that for $\omega < 2\Delta_1$, the 2-band contribution to the optical conductivity at T=0 should be
\begin{equation}
{\Re e}\ \sigma_{2 band}(\omega) = \frac{2 e^2(v_{D1} + v_{D2})^2A\xi\sin^2({k_{F1}a})}{a^2\hbar \omega}\delta(\omega-M_1).
\end{equation}
As in principle there is only one velocity scale in the (massive) SO(8) Gross-Neveu model, $v_{1}$, one might expect to find ($v_{D1} + v_{D2}) \propto  v_{1}$. However, one might expect the opening of this gap to strongly renormalize $v_1$.

An interesting situation emerges when the Fermi surface is partially 
{\bf truncated}, i.e., when $T_2\ll T\ll T_1$: Band 2 
is still 
{\it metallic} whereas bands 1 and 3 already form a D-Mott 
{\it insulating} state. In fact, the 
umklapp term $-\frac{g_{\rho}}{(\pi a)^2}\cos(\sqrt{8\pi}\Phi_{\rho})$ is still small, because
its renormalization 
has been (completely) cutoff by thermal effects. Band 2 then behaves as a
single-chain Luttinger liquid which produces a 
prominent Drude peak of height $\sim 2e^2v_{D1}/{\hbar}\sim 2e^2u_{\rho}K_{\rho}/{\hbar}$ at $\omega=0$; from this we see that $v_{D1} = v_{2}$, the Fermi energy of the second band. Moreover, following 
Ref.~\onlinecite{Lin}, the current 
operator $\Psi_1^{\dagger}\tau^z\Psi_1$
clearly excites the mass $\sqrt{3}\Delta_1$ excitons, as well as higher energy
continuum scattering states (preformed pairs) with energy above
$2\Delta_1$.  For clarity's sake, results have 
been summarized in Fig.~1. At very high-frequency, we
approximately recover the behavior\cite{Giam} of a single-chain at 
half-filling, i.e., $\sigma
(\omega)\propto \omega^{-1}$ (see Appendix A).
\vskip1pc

{\centering
{IV. SMALL N: THE EFFECTS OF TEMPERATURE}\\}

\vskip1pc

As mentioned previously, one certainly would expect a non-trivial role to be played by the temperature of the system.
\begin{figure}[ht]
\includegraphics[scale=0.31]{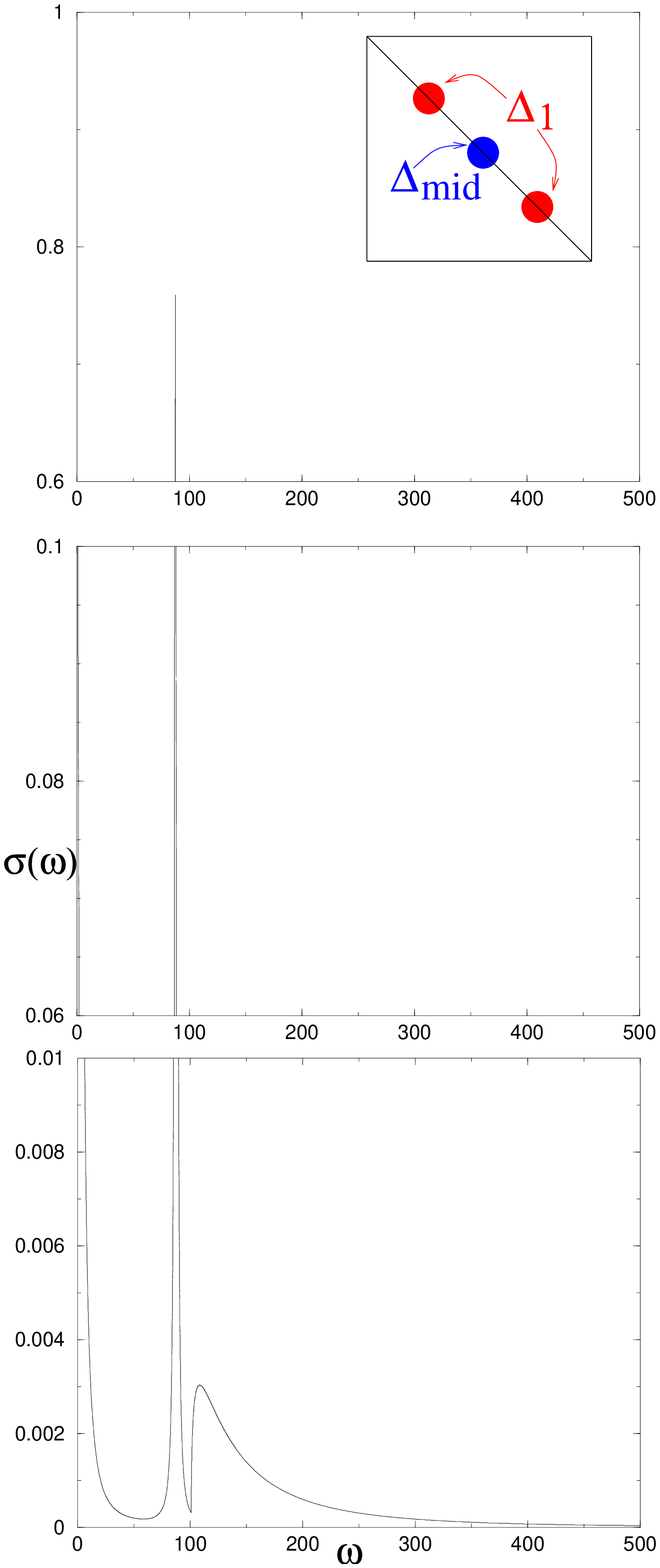}
\includegraphics[scale=0.31]{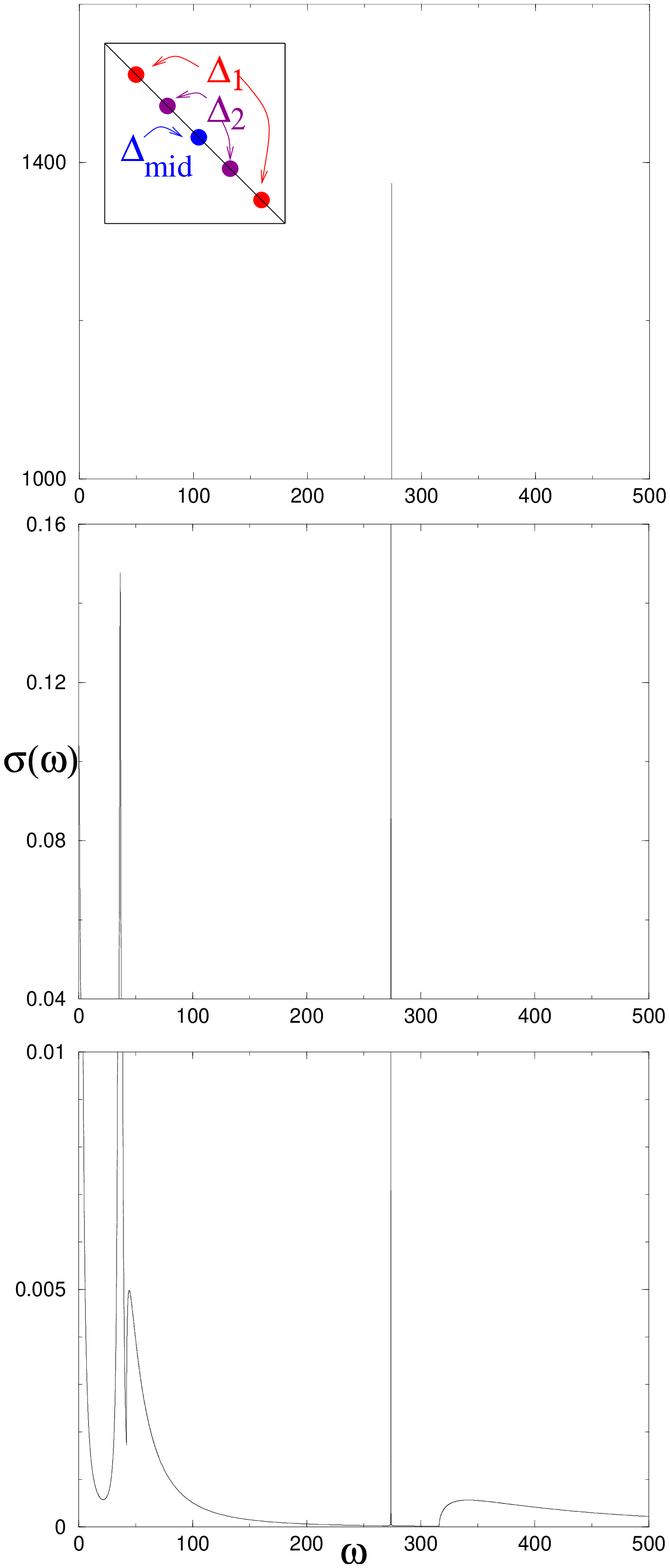}
\caption{Optical conductivity for (left) a 3-leg ladder, and (right) a 5-leg ladder.  Here we have made the assumptions: $\Delta_{mid} \approx 1 meV \approx 10 K$ and $t \approx 250 meV$, to yield $\frac{U}{t} \approx 0.36$, which defines: (left) $\Delta_1 \approx 50.4 K$ and (right) $\Delta_1 \approx 158 K$ and $\Delta_2 \approx 20.9 K$.  Further, temperature has been chosen to reflect the situation of minimal metallic conductivity: (left) $T \approx 20 K$ leading to approximately one third of the spectral weight in the Drude-like peak; and (right) $T \approx 15 K$ leading to approximately one fifth of the spectral weight in the Drude peak, and temperature broadening effects have been added in a minimal way (see text).}  
\end{figure}
While at strictly T=0, one would expect to see a $\delta$-function peak arising from the bound-state excitonic contribution to the optical conductivity, this peak should be thermally broadened.  At $T=0$, a mean field treatment would predict the 
spectral weight in the high-energy
continuum to have a square-root singularity reflecting the van Hove singularity at the bottom of a band\cite{Lin}; 
At $T\neq 0$, the scattering time between the fermions
$\Psi_1$ becomes finite and one
would expect $\sigma(\omega=2\Delta_1,T)\approx e^{\Delta_1/T}$ (finite).  On the other hand, since the SO(8) Gross-Neveu model is integrable, the optical conductivity for this state can be found exactly in principle, as it consists of the bound-state, 2 particle scattering, and higher particle scattering contributions.  Konik and Ludwig{\cite{konlud}} have calculated the (zero temperature) optical conductivity of the bound-state and the 2-particle scattering contribution of this model to be (k=0),
\begin{eqnarray}
&&\hskip-1pc{\Re e}\ \sigma_{2 band}(\omega)= A^2\hbox{\Large{(}}\delta(\omega - \sqrt{3}m)\frac{2\Gamma(1/6)}{9m^2\Gamma(2/3)}\sqrt{\frac{\pi}{3}}\nonumber\\&& \exp\left[-2\int_0^{\infty}\frac{dx}{x} \frac{G_c(x) \sinh^2(x/3)}{\sinh(x)}\right]\nonumber\\&& + \theta(\omega - 4 m^2) \frac{12 m^2}{(\omega^2-3m^2)^2}\frac{\sqrt{\omega^2 - 4m^2}}{\omega}\exp\hbox{\Large{[}}\nonumber\int_0^{\infty}\frac{dx}{x}\\&&\times\frac{G_c(x) (1-\cosh(x)\cos(\frac{x\cosh^{-1}\left[\frac{\omega - 2 m^2}{2 m^2}\right]}{\pi}))}{\sinh(x)}\hbox{\Large{]}}\hbox{\Large{)}},  
\end{eqnarray}
where A is an overall constant, m=$\Delta_1$ is the mass of an elementary SO(8) Gross-Neveu particle, and $G_c = 2\frac{\cosh(x/6) - \sinh(x/6)e^{-2x/3}}{\cosh(x/2)}$.  In particular, it is useful to notice that the square-root singularity predicted by mean field theory does not survive in the exact calculation. 
  Integration of their result,
 (assuming higher order scattering processes do not dominate) shows that an astounding 69\% of the spectral weight of the two-leg ladder (at zero temperature) is expected to lie within the bound-state.    
 As noted, their results are exact up to the onset of the three-particle continuum, provided one stays at $\omega < \omega_c$, where $\omega_c$ is the temperature at which the RG couplings reach their approximate SO(8) coupling ratios.  Above this frequency, presumably the chains will begin to behave more and more as individual chains, such that one might expect a crossover from the $\frac{1}{\omega^3}$ high frequency behavior found by Konig and Ludwig{\cite{konlud}} to an approximate $\frac{1}{\omega}$ as found by Giamarchi{\cite{Giam}}.  The effects of such a crossover have been considered in Appendix A.

In the hopes of making a more simple comparison to experimental curves, we here make the simplest assumptions consistent with a low temperature broadening of the bound-state Gross-Neveu peaks following the treatment by Lin et al{\cite{Lin2}}. There, it was argued that for temperatures T$<<$m, one could essentially solve the classical Boltzmann equation to obtain a particle density due to thermal excitations of $n \sim \sqrt{mT} e^{-\frac{m}{T}}$.  It was further argued that this approximation was self-consistent as the mean free path resulting $l\sim \frac{1}{n}$ would be much greater than the de Broglie wavelength $\lambda \sim \frac{1}{\sqrt{mT}}$, meaning that although the scattering processes themselves were quantum-mechanical, one could think of the particles essentially classically in between these time periods, leading to a time between collisions $\tau\sim \frac{2\pi e^{\frac{m}{T}}}{T}$.  Furthermore, they argued that the shape of temperature-broadened bound-state peak should be proportional to this single-particle scattering time, additionally satisfying a scaling ansatz, $\sigma_{peak} \sim\tau\Sigma((\omega-\sqrt{3}m)\tau)$ for $(\omega-\sqrt{3}m)<<m$.  

For our (N-leg) truncated ladders then, the simplest ansatz might be to assume that each $\delta$-function peak is broadened by the lifetime of the particles associated with each successive gap on the Fermi surface; that is, to make the blanket assumption $\tau_i \sim \frac{2\pi e^{\frac{\Delta_i}{T}}}{T}$ and represent the temperature broadened exciton peaks as $\sigma_{peak} \propto \frac{1}{\pi} \frac{\tau_i}{1 + ((\omega - \sqrt{3} \Delta_i) \tau_i)^2}$.  In the case of interest, when the temperature is above one of the gaps, one would expect the spectral weight to be distributed about zero, but the breadth of the peak might still be governed by the ratio of the energy scales of the gap relative to the temperature.  In such a way, we would expect to simply shift the peak to zero, but effectively keep the same form as the excitonic peak broadenings.  To create Fig. 2, we have taken this ansatz in combination with the exact results of Konig and Ludwig{\cite{konlud}}, and the assumption that as the ladder separates into effective 2- and 1-band ladder contributions, the spectral weight in each should be conserved.  Additionally, we have input experimental parameters in such a way as might be relevant for a discussion of (TMTSF)$_2$PF$_6$, although this leads to an unreasonably high value of $\frac{U}{t}\approx 0.36$ for the validity of the SO(8) as noted in Appendix A.  We note that these results, even if one was well within the validity of the SO(8) symmetry, certainly underestimate the spectral weight in the continuum, as 3- and higher- particle processes have been neglected, in addition to the (unphysical) implicit assumption that there has not been a transfer of spectral weight between the bound-state and continuum as a function of temperature.

None-the-less, the result is a semi-quantitative picture of what one might expect to find in the optical conductivity of the ideal Hubbard ladder system at intermediate temperatures at half-filling (ie. if one could make measurements at frequencies sufficiently small frequencies and temperatures).  The percentage of spectral weight in the Drude-peak would be expected to be diminished slightly as one would expect a small Mott contribution close to the relevant energy gap (here 1 meV), but the total spectral weight of this feature when one has only one gap above the temperature scale would be expected to be of the order $\frac{1}{N}$, with a progressive narrowing of the width of this feature as the temperature decreased (as seen for example in (TMTSF)$_2$PF$_6$ by Vescoli et al{\cite{Vescoli}}.  For the small number of legs where this truncation should occur, one should not be able to achieve such a small percentage of the total weight (1\%){\cite{Vescoli}} as seen experimentally{\cite{Vescoli}}.   
\vskip1pc

{\centering
{V. LARGE N: ANTIFERROMAGNETIC 4-BAND HOLE PAIRS }\\}

\vskip1pc

Let us now increase considerably the number of chains (we
consider the solvable 
case where the energy difference between band pairs $\sim 1/N$ is 
larger than the relevant gaps). As long as ``4-band'' 
interactions between bands $(i,k,\bar{\imath},\bar{k})$ 
$(k_{Fi}\rightarrow k_{Fk},k_{F\bar{\imath}}\rightarrow k_{F\bar{k}})$ 
--which stand for 2D-like antiferromagnetic interactions (Fig.~3)-- 
remain {\it small}, our system is a bunch
of (almost) decoupled bands and is still 1D-like. {\it 
However, at low-energy, the system clearly behaves as in 2D 
(See Ref.~\onlinecite{Urs})!} Indeed, two-band and single-band 
interactions will be dominantly

\begin{figure}[ht]
\centerline{\epsfig{file=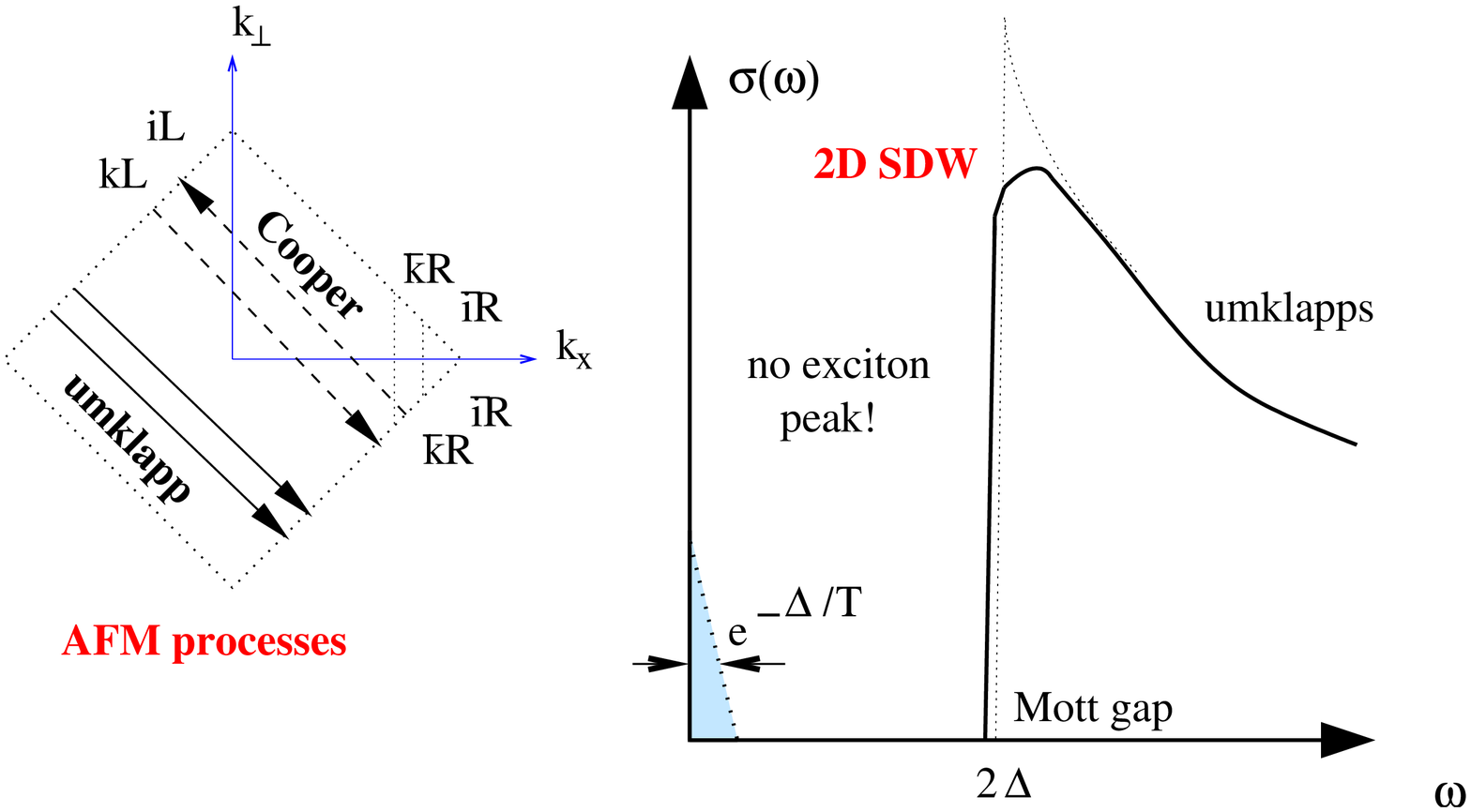,angle=0.0,height=4.5cm,width=
8.75cm}}
\caption{Optical conductivity of the N-band model for large (but finite) N
and $T<T_{sdw}$.
2D antiferromagnetic (AFM) interactions break the SO(8)
symmetry (exciton peaks {\it vanish}); The system flows to a 2D SDW 
state with a uniform charge gap.}
\end{figure}

\hskip -0.3cm renormalized by the ``four-band'' interactions:
Certain couplings of the band
pairs $(i,\bar{\imath})$ still grow and tend to approach fixed ratios, however
these differ from the small N ratios.
Furthermore, for bands $k$ and $i$ which are close together on the FS, 
as $k\rightarrow i$, the four-band couplings
become the same as the corresponding two-band couplings. 
The one-loop RGEs and theoretical details 
are given explicitly in Ref.~\onlinecite{Urs}.
To summarize, all 
the
 band pairs now {\it strongly} 
interact on the 2D FS, and this leads to a {\bf unique} 
single-particle (Mott) gap $\Delta$ which can be calculated with the 
asymptotic ratios.
Using Eq. (\ref{sdw}) for $t_{\perp}\rightarrow t$, we find
\begin{equation}
\Delta\sim T_{sdw}=t e^{-\alpha t/(U\ln N)}.
\end{equation}
The logarithmic 
corrections come from the fact that $v_1\approx ta/N$, a precursor of
van Hove singularities in 2D.  Let us repeat that here
$\ln N<t/U$ for the (strict) validity of our calculations. 
In this limit, it is yet
possible to derive the ground-state properties using bosonization\cite{Urs}.

The pinning of the antisymmetric spin mode between bands 
$i$ and $\bar{\imath}$, 
$\Phi_{\sigma i\bar{\imath}-}\approx0$, produces
spinon confinement leaving as physical particles spin-1 magnons. In the symmetric spin sector, only the differences 
$\Theta_{\sigma i\bar{\imath}+}-\Theta_{\sigma k\bar{k}+}$ are fixed, such that
the total magnon mode(s), given by $\Theta_S=\sqrt{2/N}\sum_{i=1}^{N/2} 
\Theta_{\sigma i\bar{\imath}+}$ and $\Phi_S=\sqrt{2/N}\sum_{i=1}^{N/2} 
\Phi_{\sigma i\bar{\imath}+}$, remain {\it gapless}. The spin system behaves 
increasingly as a 2D
SDW in agreement with the strong $U$ limit\cite{Heinz2}, and 2D AFM 
interactions explicitly break the SO(8) symmetry
of individual band pairs as it becomes more favorable to pin the mode
$\Theta_{\sigma i\bar{\imath}+}-\Theta_{\sigma k\bar{k}+}$ rather than
$\Phi_{\sigma i\bar{\imath}+}$\cite{Urs}. Finally, 
2D AFM interactions do not much affect 
the ground-state charge properties of band pairs, i.e., as in the small-N limit $\Phi_{\rho i\bar{\imath}+}\approx0$ (symmetric charge mode) and 
$\Theta_{\rho i\bar{\imath}-}\approx 0$ (antisymmetric superfluid mode).

 Let us 
now discuss the optical conductivity in
this large-N limit.  First, one can easily show that exciton peaks (that were a blatant 
signature of the underlying SO(8) symmetry for small-N) cannot persist.  Integrating out the spin sector -which now produces antiferromagnetism and 
then totally {\it decouples} from the gapped charge sector at low energy-
the symmetric charge mode $\Phi_{\rho i\bar{\imath}+}$ is simply
described by a Sine-Gordon model\cite{Urs} with 
$\beta\approx\sqrt{4\pi}$. The spectrum now contains only  
solitons and anti-solitons (bound hole- (electron) pairs)
with dispersion $\epsilon(p)=\sqrt{p^2+\Delta^2}$. At $\omega\approx 0$ 
and $T\ll \Delta$, the density of 
excited carriers evolves like $n\approx 
e^{-\Delta/T}$, which implies that the Drude peak has an
exponentially vanishing weight, reflecting the simultaneous opening of a 
Mott gap on the whole 2D FS. Finally, at high-frequency, we still find
$\sigma(\omega)\propto \omega^{-1}$ (we have almost decoupled bands).

\vskip1pc

{\centering
{VI. CONCLUSIONS }\\}

\vskip1pc

In closing, we
are able to provide a microscopic (low energy) theory for half-filled
N-leg Hubbard ladders 
--even when N is very large-- taking into 
account both the ubiquity of Mott physics and the emergence of prominent 
antiferromagnetism.  At small N, we have demonstrated that it is at least 
possible within the Hubbard model at half-filling for a system to exhibit 
metallic behavior over a finite temperature range while maintaining a large 
Mott contribution, a situation not unlike that seen in 
(TMTSF)$_2$PF$_6$--although the T$^2$ resistivity encountered in that case 
has not been recovered here, the relative spectral weight of the Drude feature would be ${\mathcal{O}}(\frac{1}{N})$ rather than 1\%, and the relevant energy scales for this system are well outside the range of validity of our method.  At high frequencies one should be able to employ a memory function approach to extract a detailed frequency dependence of the optical conductivity even in regions of moderate coupling.   For 
large N (and $U<t/\ln N,t_{\perp}/\ln N$), the Mott gap opens simultaneously
on the 2D FS and, as a precursor
of superconductivity, charge excitations
resemble hard-core bosons (preformed pairs), whereas 
spin excitations are 
gapless magnons (bosons). The optical conductivity is mostly sensitive 
to the Mott gap opening 
on the 2D FS (and not to the long-range magnetism) 
and excitons cannot persist! This seems to
be in agreement with results on {\it half-filled} cuprates\cite{optical}.
When doping, we stress the
occurrence of a spin gap and the progressive growth of d-wave
phase coherence on the 2D FS.
In order to properly discuss the pseudogap phase of 
cuprates, 
we must include the effect of 
a next nearest neighbor hopping\cite{John} $t'$.  It is conceivable that such a term might provide a mechanism for the creation of the ``mid-gap'' states of the doped cuprates{\cite{arima,optical}} (which bear a qualitative resemblance to the two-leg ladder contribution as outlined in Fig. 2), if the effective value of U were to vary substantially around the Fermi surface.

We acknowledge discussions with C. Bourbonnais, R. Duprat, 
T. Giamarchi, S. Kancharla, A.-M. Tremblay, 
A. Tsvelik, and financial support by FQRNT and NSERC.

\vskip1pc

{\centering
{APPENDIX A:  GENERIC FEATURES FROM ONE TO THREE CHAINS  }\\}

\vskip1pc

A serious problem for this theory is the weakness of U required for the theory to flow to the strong coupling SO(8) fixed point. 
 In particular, this imposes that our gap sizes and hence all of our temperature scales are outrageously low, as the price to pay for the rigor of our treatment.  In particular, to actually reach SO(8) within the validity of the 1 loop RG, the half-filled two-leg ladder requires $\frac{U}{t}\leq 2\times 10^{-6}$, yielding a gap $\sim te^{-500000}$, which for all intents and purposes experimentally might as well be zero.  This can be seen from Fig.4, an explicit plot of the ratios of the couplings for the two-leg ladder as a function of the normalized number of steps.  For a definition of the couplings of the two and three leg ladder, the reader is referred to Ref.11 and Ref.22.  One sees that the curves described by the 1-loop RG equations are universal, although the region of validity decreases as U increases.  In the inset, this is seen as a function of the maximal coupling, with ratios expected to reach 1 and 0 if the SO(8) fixed point is reached.  
\begin{figure}[ht]
\includegraphics[scale=0.55]{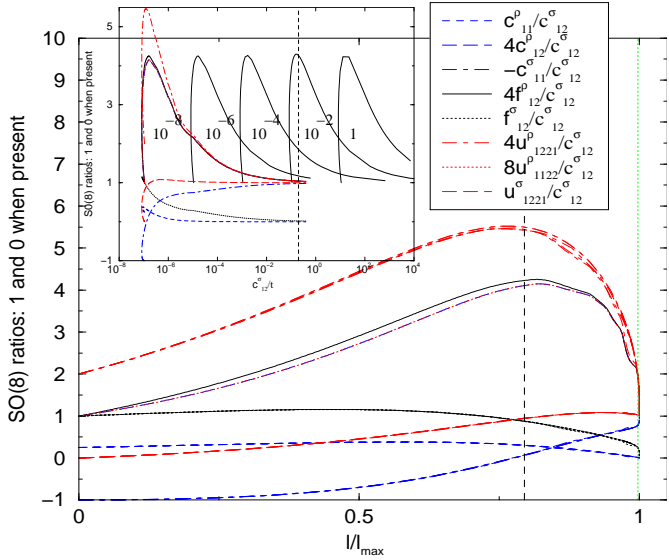}
\caption{Validity of the 1-loop RG: (main) Scaling of the RG equations as a function of the number of steps, normalized by the point at which approximate SO(8) symmetry is found (l$_{max}$).  For a definition of the couplings of the two and three leg ladder, the reader is referred to Ref.11 and Ref.22.   c$^{\rho}_{12}$, f$^{\rho}_{12}$, and u$^{\rho}_{1122}$ have been plotted for U=10$^{-8}$ due to their close proximity, while the remaining curves have been plotted for t=0.09, U=$\{10^{-8},10^{-6},10^{-4},10^{-2},1\}$ with l$_{max}$ = (50993400, 508230, 5118.84, 51.7196, 0.51743) respectively.  The vertical black dashed line delineates the point where the 1-loop RG equations are expected to break down for U=10$^{-2}$, the vertical gray dotted line for U=10$^{-4}$.  (inset)  The same curves plotted as a function of the ratio of the coupling $\frac{c^{\sigma}_{12}}{t}$. The vertical dashed line at 0.2, represents the point beyond which the 1-loop RG should not be valid.  In principle, these curves could be used to extract the frequency dependence of the umklapp contributions to the two- (and three-) leg ladders as sketched in Fig. 7 and Fig. 8.}  
\end{figure}

Even in the absence of SO(8) symmetry, one clearly can say something about the evolution of the couplings (until the higher order contributions in U become relevant), which means that in principle one could calculate the $\omega$-dependence of the optical conductivity at high frequencies for the two and three leg ladders using a memory function approach{\cite{mem}} akin to Giamarchi's work on the single chain.{\cite{Giam}}
In this spirit, here we provide a heuristic sketch of such a derivation for the 3-leg ladder, and what one would expect to be the region of its applicability of this perturbative approach.  At very high frequencies, one would expect the optical conductivity to break into two components roughly as,
\begin{equation}
\sigma(\omega) = i\frac{2e^2u_{\rho}K_{\rho}}{\hbar(\omega + M_{1 band}(\omega))} + i\frac{4e^2u_{\rho+}K_{\rho+}}{\hbar(\omega + M_{2 band}(\omega))},
\end{equation} 
where the memory functions, $M_{1 band}(\omega)$ and $M_{2 band}(\omega)$ can be computed perturbatively in linear response theory as outlined in Ref. 19 and 24 respectively.  In particular one finds that{\cite{Giam}} in the limit $\omega \gg T$,
\begin{equation}
M_{1 band} = \kappa g_{\rho}^2\Gamma^2(1-2K_{\rho})\sin(2\pi K_{\rho}) e^{-i\pi (2K_{\rho}-1)}\omega^{4K_{\rho}-3},
\end{equation}
where $\kappa \approx \frac{K_{\rho}}{\pi^3 a^2}\left(\frac{\pi a}{u_{\rho}}\right)^{4K_{\rho}-2}$, and
\begin{eqnarray}
M_{2 band} &=& G_{\chi} \sum_{\chi=1}^4\hbox{\Large{(}}(g_{\chi})^2 \Gamma^2(1-(K_{\chi})) \sin(\pi(K_{\chi})) \times \nonumber \\ && \hskip2pc e^{-i\pi (K_{\chi}-1)} \omega^{2(K_{\chi}) - 3}\hbox{\Large{)}},
\end{eqnarray}
where $G_{\chi} \approx \frac{K_{\rho+}\sin^2(k_{F_1}a)}{4^3 \pi^3 a^2}\left(\frac{\pi a}{u_{\rho+}}\right)^{2K_{\chi}-2}$
\begin{displaymath}
g_{\chi} = \left\{ \begin{array}{ll}
2 u^{\sigma}_{1331} & \hskip2pc \textrm{$\chi$ = 1}\\
u^{\sigma}_{1331} + 4 u^{\rho}_{1331}  &\hskip2pc \textrm{$\chi$ = 2}\\
u^{\sigma}_{1331} - 4 u^{\rho}_{1331} &\hskip2pc \textrm{$\chi$ = 3}\\
16 u^{\rho}_{1133} &\hskip2pc \textrm{$\chi$ = 4,}
\end{array} \right.
\end{displaymath} 
and
\begin{displaymath}
K_{\chi} = \left\{ \begin{array}{ll}
K_{\sigma+} + K_{\rho+} & \hskip2pc \textrm{$\chi$ = 1}\\
K_{\sigma-} + K_{\rho+}  &\hskip2pc \textrm{$\chi$ = 2}\\
\frac{1}{K_{\sigma-}} + K_{\rho+} &\hskip2pc \textrm{$\chi$ = 3}\\
\frac{1}{K_{\rho-}} + K_{\rho+} &\hskip2pc \textrm{$\chi$ = 4.}
\end{array} \right.
\end{displaymath}
For the three-leg ladder, the Luttinger coefficients can be written as, K$_{\rho} = \sqrt{\frac{\pi v_2 - c_{22}^{\rho}}{\pi v_2 + c_{22}^{\rho}}}$, K$_{\rho+}=\sqrt{\frac{\pi v_1 - c_{11}^{\rho} - f_{13}^{\rho}}{\pi v_1 + c_{11}^{\rho} + f_{13}^{\rho}}}$,  K$_{\rho-}=\sqrt{\frac{\pi v_1 - c_{11}^{\rho} + f_{13}^{\rho}}{\pi v_1 + c_{11}^{\rho} - f_{13}^{\rho}}}$, K$_{\sigma+} = \sqrt{\frac{\pi v_1 + \frac{c_{11}^{\sigma}}{4} + \frac{f^{\sigma}_{13}}{4}}{\pi v_1 - \frac{c_{11}^{\sigma}}{4} - \frac{f^{\sigma}_{13}}{4}}}$, and K$_{\sigma-} = \sqrt{\frac{\pi v_1 + \frac{c_{11}^{\sigma}}{4} - \frac{f^{\sigma}_{13}}{4}}{\pi v_1 - \frac{c_{11}^{\sigma}}{4} + \frac{f^{\sigma}_{13}}{4}}}$.  In the high frequency limit, all couplings, with the exception of u$^{\sigma}_{1331}$ which vanishes so that one recovers the form of Ref. 24, are of order Ua/$\hbar$ (if one is interested in the exact values, the reader is referred to Table I of Ref. 22), such that the Luttinger coefficients all closely approach 1.  Eq. 22 is then the sum of three identical terms.  In particular, one finds that Im[M($\omega$)] $\propto$ $U^2 \omega^{2K_{\rho+} -1} \Gamma^2(-K_{\rho+})\sin^2(\pi K_{\rho+})$.   In the opposite limit, the strong coupling SO(8) fixed point corresponds to 0 $<$ g $\approx$ c$^{\sigma}_{13} \approx$ 4 f$^{\rho}_{13}$ $\approx$ 4 c$^{\rho}_{13} \approx$ 4 u$^{\rho}_{1331} \approx$ 8 u$^{\rho}_{1133} \approx$ u$^{\sigma}_{1331} \approx$ -c$^{\sigma}_{11}$, so that one sees that while K$_{\rho}$ remains close to 1, spin and charge gaps drive K$_{\rho +}$,$\frac{1}{K_{\rho-}}$, K$_{\sigma\pm}$ towards 0.  If this limit were strictly attainable, then $g_3 = 0$, and Eq. 22 again would become the sum of three identical terms, with a na{\"i}ve $\omega^{-3}$ dependence to Im[M($\omega$)], although it is essential to note both that: the coefficient of such a term would also be highly frequency dependent, and for the small frequencies required for the validity of our theory, M($\omega$) would outgrow the perturbative regime required for the memory function's linear response to be valid (despite the smallness of G) a little before this point. 
\begin{figure}[ht]
\includegraphics[scale=0.5]{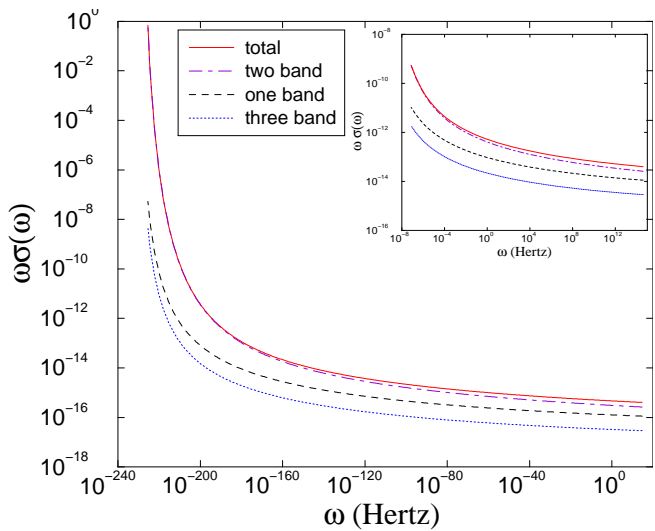}
\caption{Using the memory function, we can extract the frequency dependence of the ``high frequency'' optical conductivity (in S.I. units).  Here we have removed the $\frac{1}{\omega}$ dependence for U/t $\approx$ 0.011.  Notice that one still has substantial frequency dependence in the crossover region and that the magnitude is quickly killed at high frequencies as required to satisfy the optical conductivity sum rules.  The asymptotic high frequency dependence is $\sigma(\omega) \sim \omega^{-1.006}$ while as one approaches the Mott gap this changes dramatically, in the small frequency limit here  $\sigma(\omega) \sim \omega^{-2.3}$.  In the inset, we show U/t $\approx$ 0.11, to emphasize the small frequencies are an artifact of the weak coupling limit.  The choice t=0.25 meV=$4*10^{14}$Hz has been taken, such that we are restricted to $\omega<2*10^{14}$Hz. }
\end{figure}

Unlike at the strong coupling point, in the high frequency limit, 3-band umklapp processes are also relevant, and contribute additional terms to both Eq. 21 and Eq. 22.  Letting $g_{\rho}\rightarrow \frac{g_{\nu}}{8}$,2K$_{\rho} \rightarrow$ K$_{\nu}$, K$_{\chi} \rightarrow$ K$_{\nu}$ and $g_{\rho+}\rightarrow \frac{g_{\nu}}{2}$ with,
\begin{displaymath}
g_{\nu} = \left\{ \begin{array}{ll}
2 u^{\sigma}_{1223} & \hskip2pc \textrm{$\nu$ = 1}\\
u^{\sigma}_{1223} + 4 u^{\rho}_{1223}  &\hskip2pc \textrm{$\nu$ = 2}\\
u^{\sigma}_{1223} - 4 u^{\rho}_{1223} &\hskip2pc \textrm{$\nu$ = 3}\\
8 u^{\rho}_{2213} &\hskip2pc \textrm{$\nu$ = 4,}
\end{array} \right.
\end{displaymath} 
and
\begin{displaymath}
K_{\nu} = \left\{ \begin{array}{ll}
\frac{1}{4}( K_{\sigma+} +K_{\rho+} + \frac{1}{K_{\rho-}} + \frac{1}{K_{\sigma-}} + 2 K_{\rho} + 2K_{\sigma}) & \textrm{$\nu$ = 1}\\
\frac{1}{4}(K_{\sigma-} + K_{\rho+} + \frac{1}{K_{\rho-}} + \frac{1}{K_{\sigma+}} + 2 K_{\rho} + \frac{2}{K_{\sigma}})  & \textrm{$\nu$ = 2}\\
\frac{1}{4}(K_{\sigma+} + K_{\rho+} + \frac{1}{K_{\rho-}} + \frac{1}{K_{\sigma-}} + 2 K_{\rho} + 2 K_{\sigma}) & \textrm{$\nu$ = 3}\\
\frac{1}{4}(K_{\sigma-} + K_{\rho+} + \frac{1}{K_{\rho+}} + \frac{1}{K_{\sigma-}} + 2 K_{\rho} + \frac{2}{K_{\rho}})& \textrm{$\nu$ = 4.}
\end{array} \right.
\end{displaymath}
Here, K$_{\sigma} = \sqrt{\frac{\pi v_2 + \frac{c^{\sigma}_{22}}{4}}{\pi v_2 - \frac{c^{\sigma}_{22}}{4}}}$.

The real part of the optical conductivity arises as,
\begin{eqnarray}
&&{\Re e}\ \sigma_{2 band}(\omega) = \nonumber \\ && \hskip-2pc \frac{4e^2 u_{\rho}K_{\rho}\Im m(M_{2 band}(\omega))}{\hbar((\omega + \Re e(M_{2 band}(\omega)))^2 + (\Im m(M_{2 band}(\omega)))^2)},
\end{eqnarray}
the condition $0.2 \ge \omega^{-1} |M_{2 band}(\omega)|$ approximates the region where this formula can be applied.  
On a 3-leg ladder the region about which our linearization is valid is $\sim \frac{1}{8}$ of the band-width (4t).  This gives a high energy cut-off on the frequencies we can describe for the 3-leg ladder: $\omega < \frac{t}{2}$ so that one can describe a large range of frequencies (see Fig. 5).  This constraint becomes more severe for N large--for N=16 one is restricted to $\omega<0.08t$--but is expected to still admit a region where the memory function returns $\sigma(\omega)\propto \frac{1}{\omega}$.
\newcommand{\onecolumn}{\columnwidth\textwidth}
{\onecolumn

\begin{figure}[ht]
\includegraphics[scale=1.0]{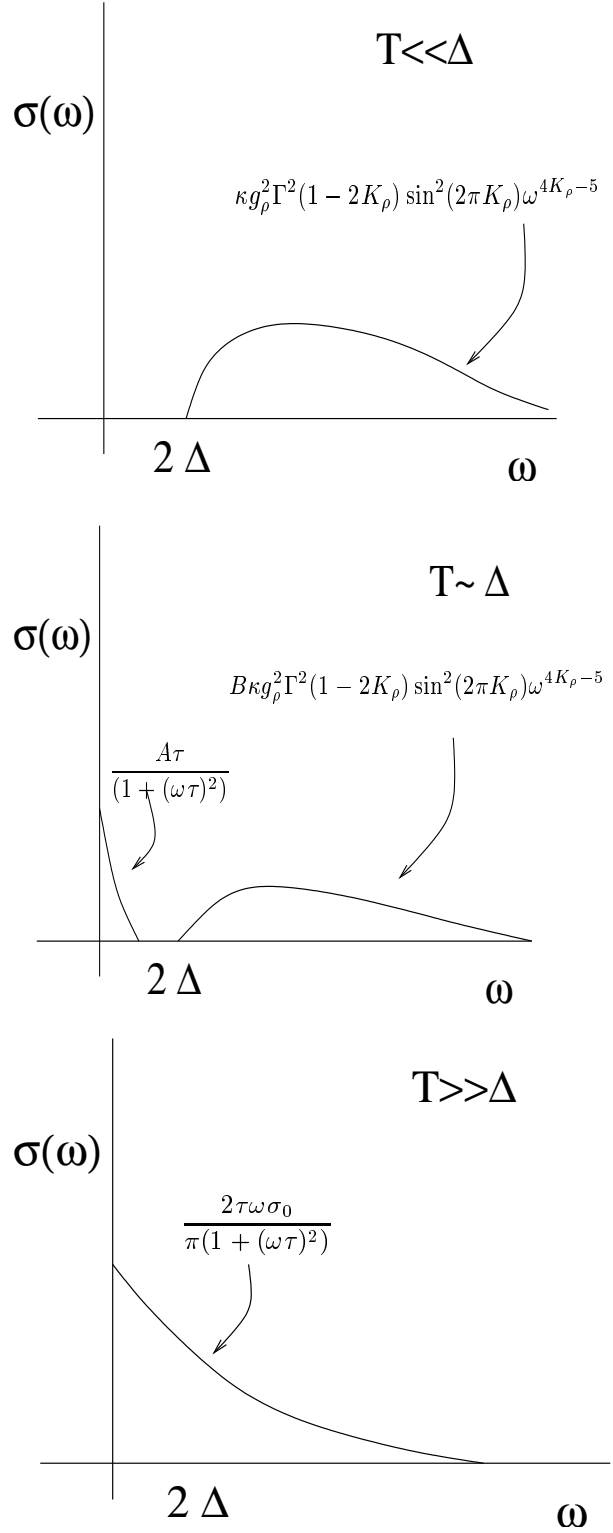}
\caption{The effects of temperature on a half-filled single chain system in weak coupling.  Here, $\tau = 2\pi e^{\Delta/T}/T$, the $1/\omega$ term is cut off at high frequencies by K$_{\rho} \rightarrow 1$, where K$_{\rho}$ and u$_{\rho}$ are the Luttinger parameters of the single chain, and at intermediate temperatures we would expect to see a partial shift of weight from above the gap to form an $\omega = 0$ Drude peak.  A, B and c are constants, g$_{\rho}$ is the strength of the the umklapp scattering as defined in Eq.9 and 21, while $\sigma_0$ is the weight 2 e$^2$u$_{\rho}$K$_{\rho}$/$\hbar$. }  
\end{figure}

\begin{figure}[ht]
\includegraphics[scale=0.75]{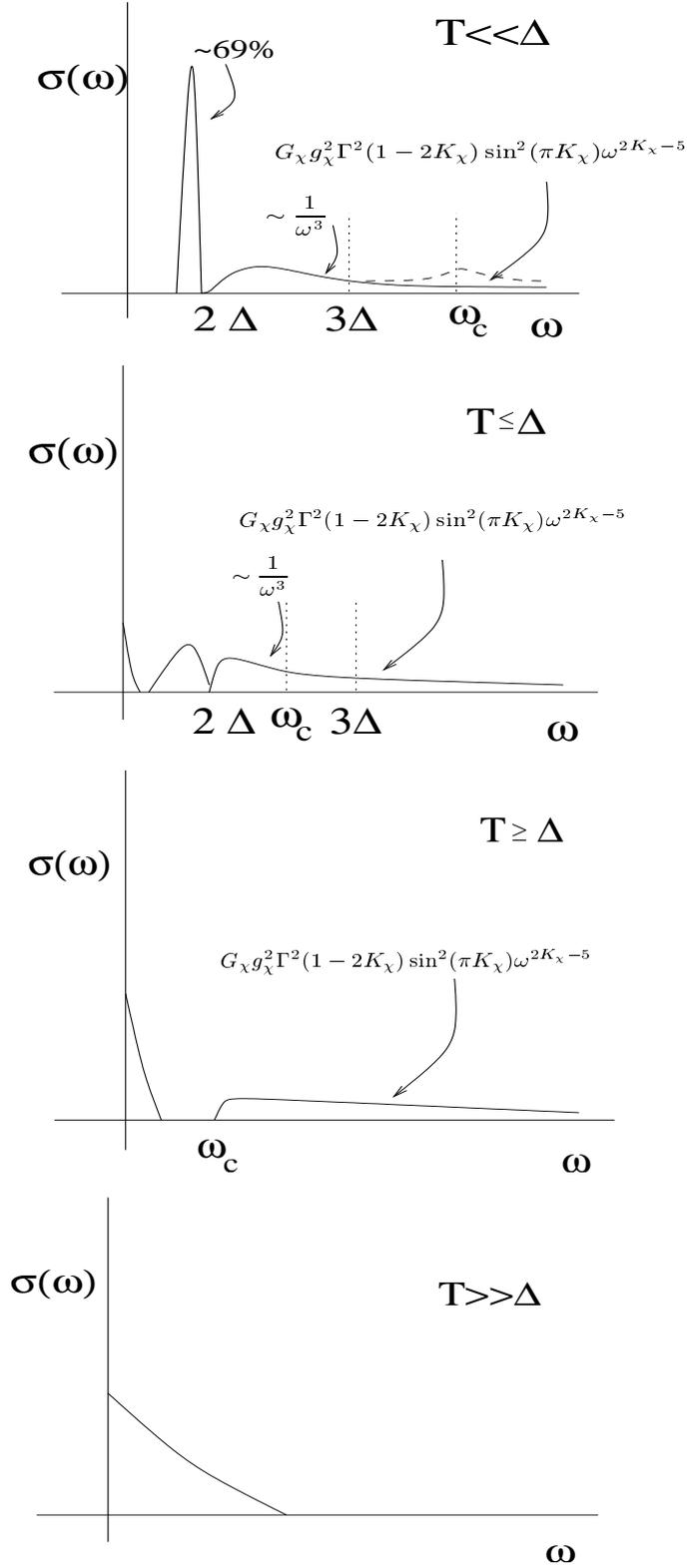}
\caption{The evolution of the two-leg ladder optical conductivity as a function of temperature, were the frequencies available experimentally.  (top) The majority of the spectral weight would lie in a sharp peak at $\sqrt{3}\Delta$; the onset of 2-particle scattering would be as $\sqrt{\omega-2\Delta}$; the decrease up to 3$\Delta$ would go as $\omega^{-3}$, and cross over to the $\omega^{-1}$ form of Giamarchi\cite{Giam}, cut off at high frequencies by the larger of the resistive umklapp scattering terms, when K$_{\chi} \rightarrow 2$, (where K$_{\chi}$ is defined in the text), such that the optical conductivity sum rule is satisfied.  Note that this need not be a simple cross-over of exponents, one might find that this asymptotic behavior arrives higher than the exact low frequency result, leading to a second bump at higher frequencies as outlined with dashed lines.  Less weight would be attributed to the exciton as the temperature approached that of the gap, with increasing weight at $\omega = 0$.  }  
\end{figure}
\begin{figure}[ht]
\includegraphics[scale=0.75]{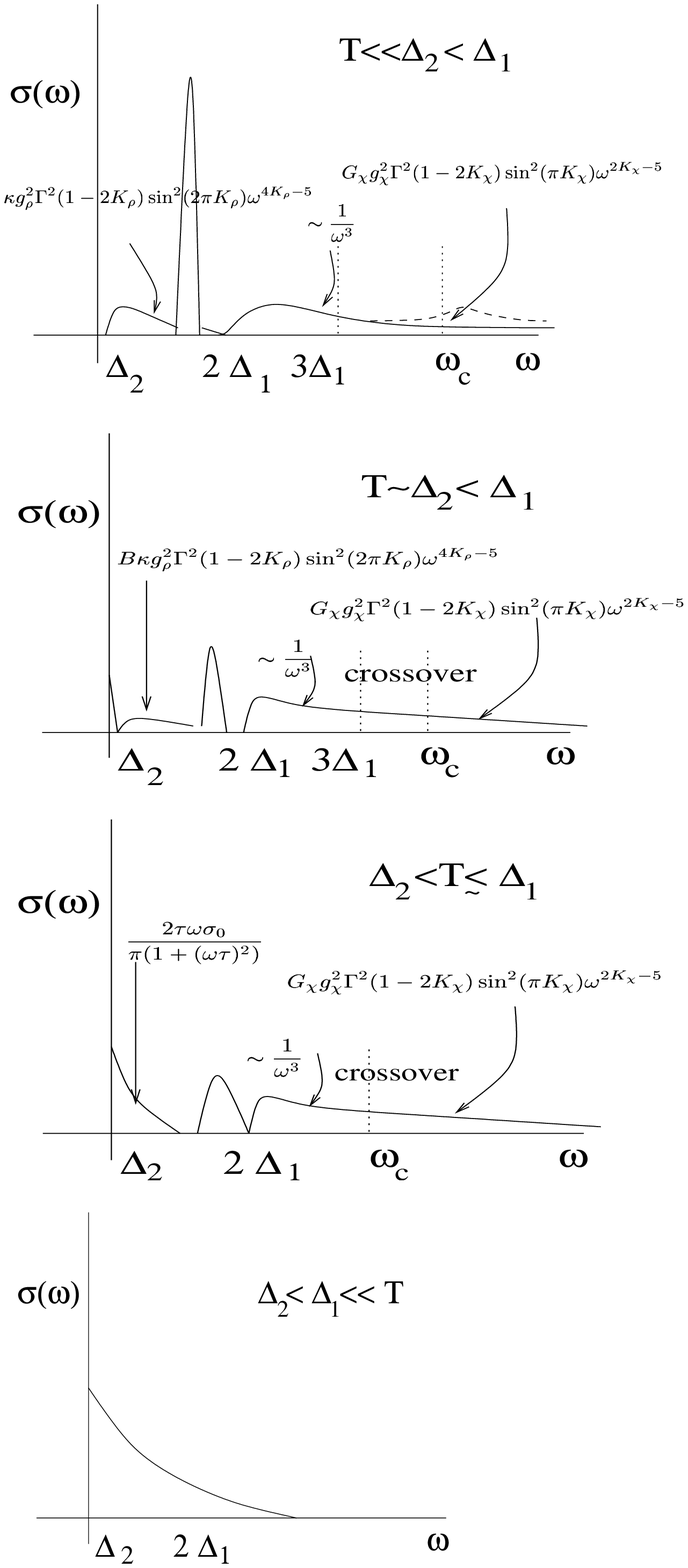}
\caption{The evolution of the three-leg ladder as a function of temperature. } 
\end{figure}}


\begin{thebibliography}{99} 


\bibitem{Claude1}
For a recent review, C. Bourbonnais in High Magnetic Fields, Applications in Condensed Matter Physics and Spectroscopy, Lecture Notes in Physics, Vol. 595, Ed. C. Berthier, L. P. Levy and G. Martinez, ISBN: 3-540-43979-X, Springer (2003).

\bibitem{Thierry}
T. Giamarchi, Physica B {\bf 230-232}, 975 (1997).

\bibitem{Tsvelik}
D. Controzzi, F. H. L. Essler, and A. M. Tsvelik, Phys. Rev. Lett. {\bf 86}, 
680 (2001).



\bibitem{Karyn1}
K. Le Hur, Phys. Rev. B {\bf 63}, 165110 (2001).

\bibitem{jerome}D. Jerome, Organic Superconductors, From (TMTSF)$_2$PF$_6$ to Fullerenes, Chap. 10, p.405, Marcel Dekker Inc. (1994).

\bibitem{Vescoli}
V. Vescoli, L. Degiorgi, W. Henderson, G. Gr{\"u}ner, K. P. Starkey, and L. K. Montgomery, Science {\bf 281}, 1181 (1998).

\bibitem{Thierry2}
S. Biermann, A. Georges, A. Lichtenstein, and T. Giamarchi, Phys. Rev. 
Lett. {\bf 87}, 276405 (2001).

\bibitem{Tsvelik2}
F. H. L. Essler and A. M. Tsvelik, Phys. Rev. B {\bf 65}, 115117 (2002).

\bibitem{phillips}
T. D. Stanescu and P. Phillips, cond-mat/0301254.

\bibitem{Maurice}
See, e.g., N. Furukawa and T. M. Rice, J. Phys. Cond. Mat. {\bf 10} L381
(1998).


\bibitem{Karyn2}
U. Ledermann, K. Le Hur, and T. M. Rice, Phys. Rev. B {\bf 62}, 16383 (2000).

\bibitem{Lin}
H. Lin, L. Balents, and M. Fisher,  Phys. Rev. B {\bf 58}, 1794 (1998).

\bibitem{Urs}
U. Ledermann,  Phys. Rev. B {\bf 64}, 235102 (2001) 
(PhD Thesis, Diss. ETH No. 14325); K. Le Hur, unpublished.

\bibitem{Zanchi}
On the N-patch model in 2D, see, e.g., D. Zanchi and H. J. Schulz, Europhys. Lett. {\bf 44}, 235 (1998).

\bibitem{optical}
S. Uchida, T. Ido, H. Takagi, T. Arima, Y. Tokura, and S. Tajima, Phys. Rev. B {\bf 43}, 7942 (1991); H. J. A. Molegraaf, C. Presura, D. van der Marel, P. H. Kes, and M. Li, Science {\bf 295}, 2239 (2002).

\bibitem{Lin2}
H. Lin, L. Balents, and M. Fisher,  Phys. Rev. B {\bf 56}, 6569 (1997).

\bibitem{Claude2}
R. Duprat and C. Bourbonnais, Eur. Phys. J. B {\bf 21}, 219 (2001).

\bibitem{Venky}
Z. Shuai, J. L. Br{\'e}das, S. K. Pati and S. Ramasesha, Phys. Rev. B {\bf{58}}, 15329 (1998); N. Tomita and K. Nasu, Phys. Rev. B {\bf{63}}, 085107 (2001); S. S. Kancharla and C. J. Bolech, Phys. Rev. B {\bf 64}, 085119 (2001).

\bibitem{Giam}
T. Giamarchi, Phys. Rev. B {\bf 44}, 2905 (1991).

\bibitem{konlud} R. Konik and A. W. Ludwig, Phys. Rev. B, {\bf{64}}, 155112 (2001).
\bibitem{Heinz2}
H. J. Schulz, in {\it Correlated Fermions and Transport in Mesoscopic 
Systems}, 
ed. T. Martin, G. Montambaux, J. Tran Thanh Van (Editions Frontieres, 
Gif--sur--Yvette, 1996), p. 81.



\bibitem{John}
J. Hopkinson and K. Le Hur, cond-mat/0308504.

\bibitem{mem} W. G{\"o}tze and P. W{\"o}lfe, Phys. Rev. B {\bf{6}}, 11226 (1972).

\bibitem{eugene} P. Byrne, E. H. Kim and C. Kallin, Phys. Rev. B {\bf{66}}, 165433 (2002).

\bibitem{arima} T. Arima, Y. Tokura and S. Uchida, Phys. Rev. B {\bf{48}}, 6597 (1993).
\end{thebibliography}
\end{document}